\documentclass[aps,prd,showpacs,preprintnumbers,twocolumn]{revtex4}
\usepackage{graphicx}
\usepackage{epsfig}
\usepackage{amsmath}
\usepackage{subfigure}
\usepackage{ulem}
\usepackage{color}
\newcommand{\al}{\alpha}

\newcommand{\Ga}{\Gamma}

\renewcommand{\th}{\theta}   

\newcommand{\beq}{\begin{equation}}
\newcommand{\eeq}{\end{equation}}
\newcommand{\ba}{\begin{array}}
\newcommand{\bea}{\begin{eqnarray}}
\newcommand{\ea}{\end{array}}
\newcommand{\eea}{\end{eqnarray}}
\newcommand{\bi}{\begin{itemize}}  
\newcommand{\ei}{\end{itemize}}
\newcommand{\ben}{\begin{enumerate}} 
\newcommand{\een}{\end{enumerate}}
\newcommand{\bc}{\begin{center}}
\newcommand{\ec}{\end{center}}
\newcommand{\bl}{\begin{flushleft}}
\newcommand{\el}{\end{flushleft}}
\newcommand{\br}{\begin{flushright}}
\newcommand{\er}{\end{flushright}}

\newcommand{\MeV}{{\rm MeV}}

\renewcommand{\l}{\left}
\renewcommand{\r}{\right}

\newcommand\comment[1]{ \hbox{[{\it Comment suppressed here.}\/]} }
\newcommand\hide[1]{}
\newcommand{\skipover}[1]{}

\begin{document}

\title{Inverse magnetic catalysis induced by sphalerons}

\author{Jingyi Chao$^{1}$}
\email{chaojy@mail.ihep.ac.cn}
\author{Pengcheng Chu$^{2,1}$}
\email{chupc@mail.ihep.ac.cn}
\author{Mei Huang$^{1,3}$}
\email{huangm@mail.ihep.ac.cn}
 \affiliation{$^{1}$ Institute of
High Energy Physics, Chinese Academy of Sciences, Beijing, China}
\affiliation{$^{2}$ Shanghai Jiao Tong University, Shanghai, China}
\affiliation{$^{3}$ Theoretical Physics Center for Science
Facilities, Chinese Academy of Sciences, Beijing, China}
\date{\today }
\bigskip
\date{\today}

\begin{abstract}

The recently discovered inverse magnetic catalysis around the critical
temperature indicates that some important information is missing in our
current understanding of conventional chiral dynamics of QCD, which is
enhanced by the magnetic field. In this work, we provide a mechanism to explain
that the inverse magnetic catalysis around the critical temperature
is induced by sphalerons. At high temperatures, sphaleron transitions between
distinct classical vacua cause an asymmetry between the number of right- and
left-handed quarks due to the axial anomaly of QCD. In the presence of a
strong magnetic field, the chiral imbalance is enhanced and destroys the
right- and left-handed pairings, which naturally induces a decreasing critical
temperature of the chiral phase transition for increasing magnetic field. The
inverse magnetic catalysis at finite baryon density, and the critical end point
in the presence of a strong magnetic field is also explored in this work.

\end{abstract}


\maketitle

\section{Introduction}

Strong magnetic fields exist in various physical systems in nature.
In magnetars, magnetic dipole fields can reach the order of
$10^{14}\textmd{ G}$,
and in the early universe, the magnitude of $10^{18-23}\textmd{ G}$
of magnetic field{\bf s} may appear
during the strong and electroweak phase transitions, respectively.
In the laboratory, strong magnetic field{\bf s} of the strength of
$10^{18-20} \textmd{G}$ (equivalent to $eB\sim (0.1-1.0~{\rm GeV})^2$)
can be generated in non-central heavy ion collisions~\cite{Skokov:2009qp,Deng:2012pc}
at the Relativistic Heavy Ion Collider (RHIC) or the Large Hadron Collider (LHC).
This offers a unique opportunity to study the Quantum Chromodynamics (QCD) vacuum
structure and strong-interaction matter under hadron-scale strong magnetic field.

The QCD vacuum has a non-trivial topological structure characterized by an integer-valued
Chern-Simons number $N_{cs}$ \cite{Belavin:1975fg}. At zero and low temperatures, the
different Chern-Simons sectors are connected by quantum tunneling transitions, i.e.
the instantons. At finite temperature, the gauge configurations that change the
Chern-Simons number can also be activated thermally through sphaleron transitions.
The presence of zero modes in the spectrum of the Dirac operator under the instanton and/or sphaleron fields induces an imbalance between the number of quarks with different chirality,
and results in a violation of the $\mathcal{P}$- and $\mathcal{CP}$-symmetry~\cite{Witten:1979vv,Veneziano:1979ec,Vicari:2008jw,Schafer:1996wv}.
Indeed, the puzzle of the mass difference between $\eta$ and $\eta'$ is resolved by
the instantons induced effective $2N_{f}$-fermion interactions \cite{Schafer:1996wv}.

The possibility of local parity violation at high temperatures or in high-energy
heavy-ion collisions has  been discussed for many years \cite{Lee:19731974,Morley:1983wr,Kharzeev:1998kz}.
Recently, the heavy-ion collisions presented the conclusive
observation of charge azimuthal correlations~\cite{Abelev:2009ad,Abelev:2012pa}
possibly resulting from the anomalous Chiral Magnetic Effect (CME) \cite{Kharzeev:2007tn,Kharzeev:2007jp,Fukushima:2008xe} with
local $\mathcal{P}$- and $\mathcal{CP}$-violation. The most essential ingredient
of the CME is the chiral imbalance induced by a nonzero topological charge through the
axial anomaly of QCD, and hence an electromagnetic current can be generated along the
magnetic field, which will induce the charge separation effect.

Another important aspect of QCD is that the chiral symmetry is spontaneously
broken in the vacuum. The dynamical chiral symmetry breaking is due
to a non-vanishing quark anti-quark condensate, $\langle{\bar \psi}\psi
\rangle \simeq \mathbf{-}(250 {\rm MeV})^3$ in the vacuum, which induces
light Nambu-Goldstone particles, i.e. pions and kaons
in the hadron spectra. In the chiral limit when the current quark mass
is zero $m=0$, the chiral condensate $\langle{\bar \psi}\psi \rangle$ is the
order parameter for the chiral phase transition, which vanishes
at high temperatures and densities where chiral symmetry is restored.
In nature, quark masses are non-zero and the chiral condensate
can be regarded only
as an approximate order parameter. Lattice simulations revealed that for
physical quark masses in the case of $N_f=2+1$, the transition shows
a crossover feature~\cite{Aoki:2006we} at high temperatures.

In the presence of an external magnetic field, chiral symmetry breaking
and restoration has been investigated for many years. Since the 1990's, it has been
recognized that the magnetic field $\bm B$ plays the role of a catalyzer
of dynamical chiral symmetry breaking, i.e.~the chiral condensate increases
with $B$, which is called magnetic catalysis
\cite{Klevansky:1989vi,Klimenko:1990rh,Gusynin:1995nb}. Naturally,
it is expected that the chiral symmetry should be restored at a higher $T_c$
with increasing magnetic field. This is agreed by almost all
effective chiral models, e.g.~in the Nambu--Jona-Lasinio (NJL) model or Polyakov-loop
NJL (PNJL) model \cite{Klevansky:1989vi,Mizher:2008hf,Menezes:2009uc,Gatto:2011wc},
as well as in some earlier lattice simulations \cite{Buividovich:2009wi,D'Elia:2010nq}.
However, with the physical pion mass, the lattice group \cite{Bali:20111213}
has revealed the transition temperature to decrease as a function of the external magnetic
field. This phenomena is called inverse magnetic catalysis around $T_c$, which is
in contrast to the naive expectation and majority of previous results.

On the one hand, the magnetic catalysis at zero and low temperatures
(below $100\mathrm{MeV}$) is confirmed by all numerical simulations.
On the other hand, the inverse magnetic catalysis around $T_c$, discovered
in \cite{Bali:20111213}, indicates that some
important information is missing in our current understanding of conventional
chiral dynamics and chiral phase transition, which is especially enhanced
by magnetic field. There are some proposals to understand the puzzle of
the inverse magnetic catalysis around $T_c$, e.g.~due to the Nambu-Goldstone
neutral pion \cite{Fukushima:2012kc}, the mass gap in the large $N_c$ limit
\cite{Kojo:2012js}, or sea quarks \cite{Bruckmann:2013oba}. In this work, we
provide a different mechanism to explain the inverse magnetic catalysis around
$T_c$, which is, that it is induced by sphalerons.
At high temperatures, sphaleron transitions between distinct classical vacua
cause an asymmetry between the number of right- and left-handed quarks due to
the axial anomaly of QCD. Under strong magnetic field, the chiral imbalance is
enhanced and destroys the right- and left-handed pairings, which naturally
induces a decreasing critical temperature of the chiral phase transition with
the magnetic field.  In the following, we explain how this mechanism works.

\section{Enhanced chiral imbalance under strong magnetic field }

Instantons and sphalerons are finite-energy solutions of the Minkowskian
equations of motion in the pure gauge sector of QCD. Such process changes
the Chern-Simons number which is defined as
\begin{equation}
 \Delta N_{\mathrm{cs}} = \frac{g^2}{32\pi^2}\int d^4 x \;\text{Tr}[F_{a\mu\nu}\tilde{F}^{a\mu\nu}] \;,
\label{eq-windingnumber}
\end{equation}
where $F_{a\mu\nu}$ and $\tilde{F}^{a\mu\nu}$ denote the gauge field strength
tensor and its dual, respectively. Topological charge changing transitions
induce an asymmetry between the number of right- and left-handed quarks due
to the axial anomaly of QCD
\begin{equation}
 (N_R - N_L)_{t=+\infty} - (N_R - N_L)_{t=-\infty} = -2N_{f} \Delta N_{\mathrm{cs}} \;,
\label{eq-chiralimbalance}
\end{equation}
where $N_{f}$ is the number of quark flavors and $N_{R,L}$ are right and left-handed
quark number, respectively. As we have mentioned above, $\mathcal{P}$- and
$\mathcal{CP}$-violating processes are originated from the topological transitions
between two degenerate QCD vacua. In order to illustrate our point of this paper,
we emphasize that at finite temperatures, around $T_{c}$, although the quantum
tunneling event of an instanton is not thermally suppressed, its
interaction is reduced by a rearrangement of the instanton-anti-instanton
ensemble. However, as an unstable solution which decays from the top of
the energy barrier, sphalerons are excited at finite temperatures due to
the thermal fluctuations \cite{Arnold:1987zg,Moore:1997im}.

In order to describe the chiral imbalance, the authors
of \cite{Fukushima:2008xe} introduced an artificial chiral chemical potential $\mu_5$.
This chemical potential $\mu_5$ couples to the chiral density operator
$\mathcal{N}_5=\bar{\psi}\gamma^0\gamma^5\psi =\psi^\dagger_R \psi_R - \psi^\dagger_L \psi_L$, hence $n_5 = \langle\mathcal{N}_5\rangle\neq0$ can develop when
$\mu_5\neq0$. Here we stress that the topological density is a $\mathcal{CP}$-odd variable and, thus,
vanishes on average in the QCD $\th$-vacuum, and this remains true in the presence of an
external magnetic field, whenever how strong is it. Therefore the mechanism proposed in
this work can be considered as happening in a scenario when the $\mathcal{P}$ and $\mathcal{CP}$-symmtries violated locally. There will be some excited vacuum domains with
left-handed quarks dominant and some other domains with right-handed quark
dominant, i.e. even though the vacuum expectation value of $n_{5}=0$ but
$\langle n_5^2 \rangle \neq 0 $.
We also point out that the energy barrier of gauge configuration is lowered while applying an external strong magnetic field. It is because that magnetic field, $B$, couples
to the sphaleron magnetic moment, ${\vec m}$, where $E_{B}=E_{0}-{\vec m}\cdot {\vec B}$.
Therefore, the acting window of sphaleron has been enlarged, which starting around $100~ {\rm MeV}$. The lattice result in \cite{Buividovich:2009wi} gave a clear evidence that $\langle n_5^2 \rangle$ not only be finite but also increases with $eB$, which strongly support our arguments.

In a strong magnetic field, quarks are polarized along the direction of $\bm B$, which we
choose to be along the positive $z$ axis. This leaves only one free spatial dimension,
i.e.~the $z$-direction whose associated momentum is denoted by $p_z$. In the massless limit,
one can distinguish modes with right-handed chirality from modes with
left-handed chirality. If a $\mu_5 \bar \psi \gamma^0 \gamma^5 \psi$
is added in the Lagrangian density, the energy spectra for massless right- and
left-handed quarks take the form of \cite{Fukushima:2008xe}
\begin{equation}
 \omega_{R\pm} = \pm p_z - \mu_5, ~~ \omega_{L\pm} = \mp p_z + \mu_5.
\label{eq-dispersion-rl}
\end{equation}
Here $\pm$ represents the spin in the $z$-direction and $R$, $L$ the chirality.
It is emphasized in \cite{Fukushima:2008xe},
that $p_z$ is restricted to be positive for the $R_{+}$ and $L_{-}$ particle
modes so that the helicity is positive for $R_{+}$ and negative for $L_{-}$,
respectively, and $p_z$ is negative for the $R_{-}$ and $L_{+}$ particle
modes.

It is noticed that, if the chiral chemical potential $\mu_5$ is non-zero,
there will be an energy mismatch, $2 \mu_5$, between the $(L_{-},R_{+})$ and
$(R_{-},L_{+})$ pairing, which tends to destroy the chiral condensate
\begin{equation}
\langle \bar \psi \psi \rangle= \langle \bar \psi_L \psi_R  +  \bar \psi_R \psi_L  \rangle.
\end{equation}
$\mu_5$ here plays a role similar to the isospin asymmetry in a
color superconductor \cite{Shovkovy:2003uu}, where the isospin asymmetry
induces mismatched Fermi surfaces between pairings of quarks and
breaks the BCS pairing. Indeed, it was found in \cite{Chernodub:2011fr} that
the critical temperature of the chiral phase transition decreases when the
chiral imbalance $\mu_5$ increases.

If the chiral chemical potential is positive, some of the right-handed
particle modes will become occupied while some of the left-handed anti-particle
modes will be filled as well. A net chirality $n_5$ can be created
\begin{eqnarray}
n_{5} = \frac{\mu_{5}^{3}}{3\pi^{2}}+\frac{\mu_{5}T^{2}}{3}.
\end{eqnarray}

To estimate the behavior of $\mu_{5}$ under a strong magnetic field $B$, we
use the time evolution formula for chiral quark density $n_{5}$,
derived in \cite{Khlebnikov:1988sr},
\begin{equation}\label{eq:evo-n5}
\frac{\partial n_{5}}{\partial t}=(4N_{f})^{2}\frac{\Ga_{ss}}{T}\frac{\partial F}{\partial n_{5}}
\end{equation}
where $\Ga_{ss}$ is the spaleron diffusion rate at strong coupling and $F$
denotes the free energy. Therefore, we can replace the last derived term in
Eq.(\ref{eq:evo-n5}) by $\mu_{5}$ directly. The decay time is estimated to be
about $80/T$ for the physical value of $\al_{s}$ \cite{Moore:1997sn}. One is able to find that the decay time of sphaleron is much larger than other microscopic chiral processes and support to creat a stable chiral imbalanced scenario.
Therefore, we have
\begin{equation}
\mu_{5}=\sqrt{3}\pi\l(\frac{320N_{f}^{2}\Ga_{ss}}{T^{2}}-\frac{T^{2}}{3}\r)^{\frac{1}{2}}.
\label{eq-mu5-diffusion}
\end{equation}
There is no good method in real QCD to calculate diffusion rate for
sphalerons at strong magnetic fields. We take the strong
coupling limit behavior obtained in \cite{Basar:2012gh} through the
holographic AdS/CFT correspondence, where
\begin{equation}
\Ga_{ss}(B,T)=\frac{(g_{s}^{2}N_{c})^{2}}{384\sqrt{3}\pi^{5}}\l(eBT^{2}+15.9T^{4}\r).
\label{eq-diffusion}
\end{equation}
From Eqs.(\ref{eq-mu5-diffusion}) and (\ref{eq-diffusion}), we find that the
magnetic field enhances the chiral imbalance induced by sphaleron transitions,
and the chiral imbalance behaves as
\begin{equation}
\mu_5 \simeq  c \sqrt{eB},
\label{eq-mu5-B}
\end{equation}
under a magnetic field $B$, with $c$ being an unknown parameter.

\section{Inverse magnetic catalysis around $T_c$}
\label{sec:model}

We now analyze the chiral phase transition with a chiral imbalance induced by
sphaleron transitions under a strong magnetic field in the Nambu--Jona-Lasinio (NJL)
model, which is described by the Lagrangian density
\begin{align}
 \mathcal{L} &= \bar\psi\left(i\gamma_\mu D^\mu
  + \mu \gamma^0+ \mu_5 \gamma^0\gamma^5 \right)\psi \notag\\
 &\quad + G\left[\left(\bar\psi\psi\right)^2
  + \left(\bar\psi i \gamma^5 \bm\tau\psi\right)^2\right] \;,
\label{eq:four-fermi}
\end{align}
where $\mu$ is the quark chemical potential, and the covariant derivative embeds
the quark coupling to the external magnetic field. The thermodynamical potential
$\Omega$ in the mean-field approximation takes the form of
\cite{Fukushima:2010fe,Gatto:2011wc,Menezes:2009uc}
\begin{align}
 & \Omega =  \frac{\sigma^2}{4G}
  -N_c\sum_{f=u,d}\frac{|q_fB|}{2\pi} \sum_{s,k}\alpha_{sk}
  \int_{-\infty}^\infty \frac{d p_z}{2\pi} \, \omega_s(p) \notag \\
 &\quad - T N_c\sum_{f=u,d}\frac{|q_fB|}{2\pi} \sum_{s,k} \alpha_{sk}
  \int_{-\infty}^\infty \frac{d p_z}{2\pi} \notag\\
 &\qquad\times ( \ln [ 1 + e^{-\beta (\omega_s+\mu)}] + \ln [ 1 + e^{-(\beta \omega_s-\mu)}] ) \;.
\label{eq:O2}
\end{align}
Where $\sigma = -2 G\langle\bar\psi\psi\rangle$ is the chiral condensate, and
the effective quark mass $M=\sigma $. The quasi-particle dispersion relation is given by
$\omega_s^2 = M^2 + \bigl[ |\bm p| + s\,\mu_5 \text{sgn}(p_z) \bigr]^2$,
with $s=\pm 1$, $\bm p^2 = p_z^2 + 2|q_f B| k$ with $k$ a non-negative integer
labeling the Landau level. The spin degeneracy factor is
\begin{equation}
 \alpha_{sk} = \left\{ \begin{array}{ll}
  \delta_{s,+1} & \text{   for~~~} k=0,~ qB>0 \;,\\
  \delta_{s,-1} & \text{   for~~~} k=0,~ qB<0 \;,\\
  1 & \text{   for~~~} k\neq0 \;.
 \end{array} \right.
\end{equation}
\begin{center}
\begin{figure}[t,h!]
\includegraphics[width=7cm]{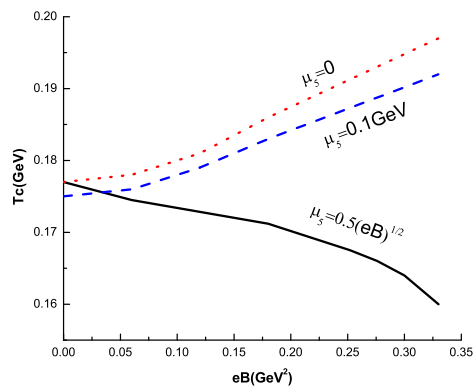}
\vspace*{-0.5cm}
\caption{ The critical temperature $T_c$ for the chiral phase
transition at $\mu=0$ as function
 of $eB$ with $\mu_5=0, 0.1 {\rm GeV}, 0.5 \sqrt{eB}$, respectively.}
\label{Fig-Tc}
\end{figure}
\end{center}
\vspace*{-1cm}
\begin{center}
\begin{figure}[t,h!]
\includegraphics[width=7 cm]{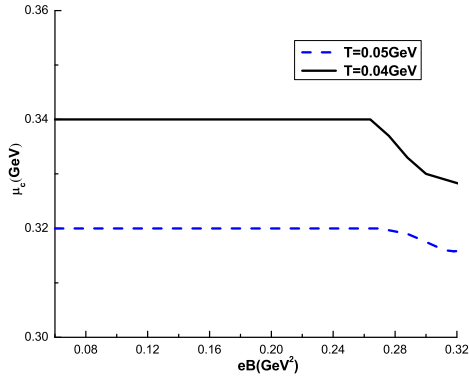}
\vspace*{-0.5cm}
\caption{The critical chemical potential $\mu_c$ in the case of
  $\mu_5=0.5 \sqrt{eB}$ as function
of $eB$ for $T=40 {\rm MeV}, 50 {\rm MeV}$, respectively. }
\label{Fig-muc}
\end{figure}
\end{center}
\vspace*{-1cm}
\begin{center}
\begin{figure}[t,h!]
\includegraphics[width=7 cm]{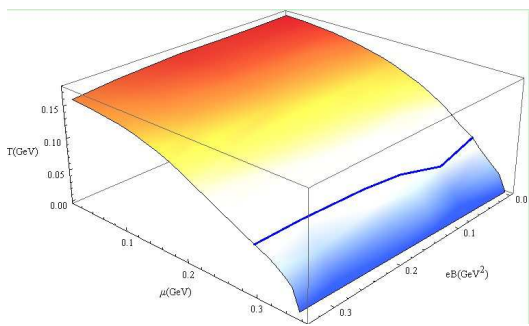}
\vspace*{-0.0cm}
\caption{The 3D chiral phase diagram, i.e. the critical temperature
  $T_c$ as function of
$\mu$ and $eB$ in the case of $\mu_5=0.5 \sqrt{eB}$. The solid line is
 for the critical end point. }
\label{Fig-3Dphasediagram}
\end{figure}
\end{center}

The model parameters are chosen as
$\Lambda = 620\MeV\;, G\Lambda^2 = 2.2\;$,
which correspond to $f_\pi = 92.4\MeV$ and the vacuum
chiral condensate $\langle\bar u u\rangle^{1/3} = -245.7\MeV$, and the
constituent quark mass $M=339\MeV$.

At zero chemical potential $\mu=0$, the critical temperature $T_c$ for the chiral phase
transition as function of $eB$ is shown in FIG. \ref{Fig-Tc} with $\mu_5=0, 0.1 {\rm GeV},
0.5 \sqrt{eB}$, respectively. For $\mu_5=0$ and $\mu_5=0.1 {\rm GeV}$, the critical temperature
$T_c$ increases with $eB$, and shows the magnetic catalysis effect. One can notice that
the critical temperature $T_c$ is reduced in the case of $\mu_5=0.1 {\rm GeV}$
compared to the case of $\mu_5=0$. If we choose the chiral imbalance $\mu_5= c \sqrt{eB}$ as in Eq.(\ref{eq-mu5-B}) with $c=0.5$, we can produce the inverse magnetic catalysis effect
as in \cite{Bali:20111213}, i.e. the critical temperature $T_c$
decreases with $eB$.

We also observe the inverse magnetic catalysis at finite chemical potential $\mu$ as shown in
FIG.\ref{Fig-muc}, which is also observed in the NJL model with $\mu_5=0$ \cite{Menezes:2009uc}
and in the holographic QCD model \cite{Preis:2010cq}. Sphalerons do not play an essential
role at low temperatures, here the inverse magnetic catalysis is induced by a finite chemical potential. In FIG. \ref{Fig-3Dphasediagram}, we show the 3-dimension chiral phase diagram,
i.e. the critical temperature $T_c$ as function
of $\mu$ and $eB$ in the case of $\mu_5=0.5 \sqrt{eB}$.
The solid line is for the critical end point (CEP). It is observed
that $(T_c, \mu_c)$ for the CEP does not change so much at high magnetic field, which is
different from the result in \cite{Menezes:2009uc} where $\mu_5=0$ and
the CEP moves toward the temperature axis with increasing magnetic field.

\section{Conclusion and discussion}

In this work, we have provided a mechanism to explain that the inverse magnetic catalysis
around the critical temperature is induced by sphalerons. At high temperatures,
sphaleron transitions between distinct classical vacua cause an asymmetry between
the number of right- and left-handed quarks due to the axial anomaly of QCD.
In the presence of a strong magnetic field, the chiral imbalance is enhanced and
destroys the right- and left-handed pairings, which naturally induces the decreasing
critical temperature of the chiral phase transition for increasing magnetic field. The CEP
under a strong magnetic field is also explored in this work, and it is found that
$(T_c, \mu_c)$ for the CEP does not change much at strong magnetic field.

At last, we emphasize that the mechanism proposed in this work can be only considered as
happening in a scenario when the $\mathcal{P}$ and $\mathcal{CP}$-symmtries violated
locally, which is supported by the lattice result in \cite{Buividovich:2009wi}.

\vskip 0.2cm
{\bf\it Acknowledgement.---} We thank valuable discussions with M. Chernodub,
D. Rischke, A.Schmitt and I. Shovkovy. This work is supported by the NSFC under
Grant No. 11275213, DFG and NSFC (CRC 110), CAS key project KJCX2-EW-N01, K.C.Wong
Education Foundation, and Youth Innovation Promotion Association of CAS.

\end{document}